\documentclass[rnote]{aa} 
\usepackage{psfig}
\usepackage{graphicx}
\usepackage[varg]{txfonts}
\usepackage[sort]{natbib}

\begin{document}

\title{A search for chaos in the optical light curve of a blazar:\\ W2R 1926+42}
\titlerunning{A search for chaos in a blazar}
\author{Rumen Bachev\inst{1} \and Banibrata Mukhopadhyay\inst{2} \and Anton Strigachev\inst{1}}
\authorrunning{Rumen, Mukhopadhyay \& Strigachev}
\institute{
$^1$ Institute of Astronomy with NAO, Bulgarian Academy of Sciences, Sofia 1784, Bulgaria,
\email{bachevr@astro.bas.bg , anton@astro.bas.bg} \\
$^2$ Department of Physics, Indian Institute of Science, Bangalore 560012, India,
\email{bm@physics.iisc.ernet.in} \\
}
\date{Accepted ... Received ...}
\abstract
{}
{In this work we search for signatures of low-dimensional chaos in the temporal behavior of the $Kepler$-field blazar W2R 1946+42.}
{We use a publicly available, $\sim$160000 points long and mostly equally spaced, light curve of W2R 1946+42.
We apply the correlation integral method to both -- real datasets and phase randomized ``surrogates".}
{We are not able to confirm the presence of low-dimensional chaos in the light curve. This result, however, still leads to
some important implications for blazar emission mechanisms, which are discussed.}
{}

\maketitle

\section{Introduction}

Within the large family of the active galactic nuclei (AGN), blazars are the
only members whose emission is produced primarily in a relativistic jet via
synchrotron and inverse-Compton processes. To be identified  as a blazar, the
jet should be oriented at a small angle to the line of sight. The blazar
spectral energy distribution consists of two peaks -- a synchrotron one,
generally peaking in the optical region, and an inverse-Compton one, peaking
in the X/$\gamma$-ray region.

Like most of the other AGN types, blazars are known to be highly variable in
the optical. In fact, they are the only AGN class, for which evidence of
significant variations clearly exists even on intra-night time scales
\citep[][and the references therein]{b1}. Blazar optical variability
(i.e. the synchrotron peak variability) on long-term time scales can generally
be described as unpredictable and lacking any periodicity
\citep[see however e.g.][]{g3}. In spite of the extensive studies throughout the years,
no clear reason for such blazar variations has been positively identified.
Unlike the radio-quiet AGN, where accretion disk instabilities are usually
invoked to account for the variations, for the blazar case similar
``variability drivers" should be searched for in the processes in a relativistic
jet. They range from purely geometrical ones (changing Doppler factor of the
emitting blob and/or microlensing from intervening foreground objects) to a
complex evolution of the relativistic particles' energy distribution (interplay
between particles acceleration and synchrotron/inverse-Compton loses).
In any case, studying blazar light curves may provide a clue to better
understand physics of the relativistic jets, provided sufficient data are
available.

A way to explore apparently random and non-periodic time series is to
search for low-dimensional deterministic chaos, which may allow to distinguish
a purely stochastic process from a process driven by a nonlinear
(and presumably chaotic)
dynamical system. While the former case most likely implies the presence of many,
acting independently factors (e.g. many independent emitting components), the
later one suggests that the entire system might be governed by a small (say 3--5)
number of degrees of freedom.

Appropriate way to search for low-dimensional chaotic signatures in a time
series is to apply the correlation integral (CI) method \citep[][see also \S 2.2]{g2}.
In the past, what concerns AGN, similar analyses have been
performed on EXOSAT X-ray light curves of several, mostly radio-quiet objects,
leading to no conclusive results \citep{l1}, due perhaps to the insufficient
number of data points, ranging from $\sim$300 to $\sim$4000 for the different
sources as well as to the relatively larger photometric uncertainty.

No conclusive results for the presence of low-dimensional chaos were also obtained
for 3C~345 \citep{p3}, based on the 800 points long optical dataset with linearly interpolated missing part.
Indications for possible fractal nature of the light curves of a few other blazars were found by \citet{m0}; however,
again -- huge missing parts of the light curves were substituted via interpolation.
Non-linear (and perhaps chaotic) variability has
been found by \citet{g1} in a 10000 points X-ray light curve of Akn~564 --
a radio-quiet AGN, observed by $Ginga$ satellite. Chaotic signature in
$\sim 20000$ points long light curves in the X-ray binary GRS~1915+105 \citep[]{misra}
was also reported.

Most of the previous work, as mentioned above, focused on Seyfert 1
galaxies, whose observed properties are very
different than those of BL Lacs (e.g. BL Lacs are very rapidly variable, highly polarized, exhibit
superluminal motion, emit up to TeV energies, while Seyfert 1-s are not) and the
physics is also believed to be different between them: beamed versus relatively isotropic emission.
In this paper, we apply the correlation integral CI method to search for low-dimensional
chaotic signatures in the optical light curve of W2R~1926+42 --- a blazar within the $Kepler$
satellite field \citep{e1}. By far, this object has the largest ever optical
dataset (consisting of almost 160000 data points in total), due to the unique
$Kepler$ capabilities to observe repeatedly the same field. More importantly, large
segments of equispaced data are available, which is specifically required
by the CI method. In this work, to the best of our knowledge, the CI analysis is
applied for the first time on such a large equidistant optical dataset of a blazar.
Fig.~1 shows the entire lightcurve of W2R~1926+42. Clearly, the object shows rapid and
significant variability over 5 times in terms of count rates, allowing a formal distinction
of "low" and "high" states, which will be treated separately later in our analysis.


\begin{figure}
\centerline{
\psfig{figure=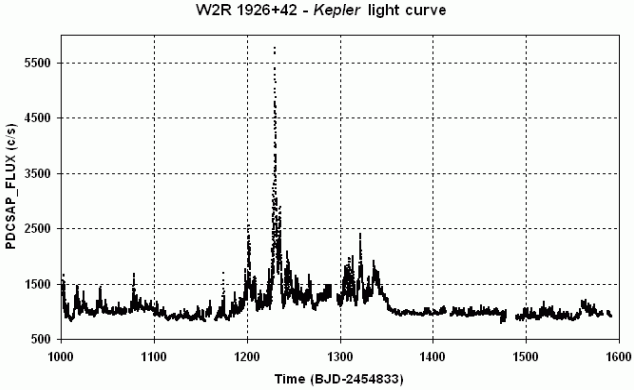,width=8.4truecm,clip=}
}
\caption[h]{
The long term light curve of W2R 1926+42 from Kepler.
}
\end{figure}

\section{Search for chaos}


\begin{figure}
\centerline{
\psfig{figure=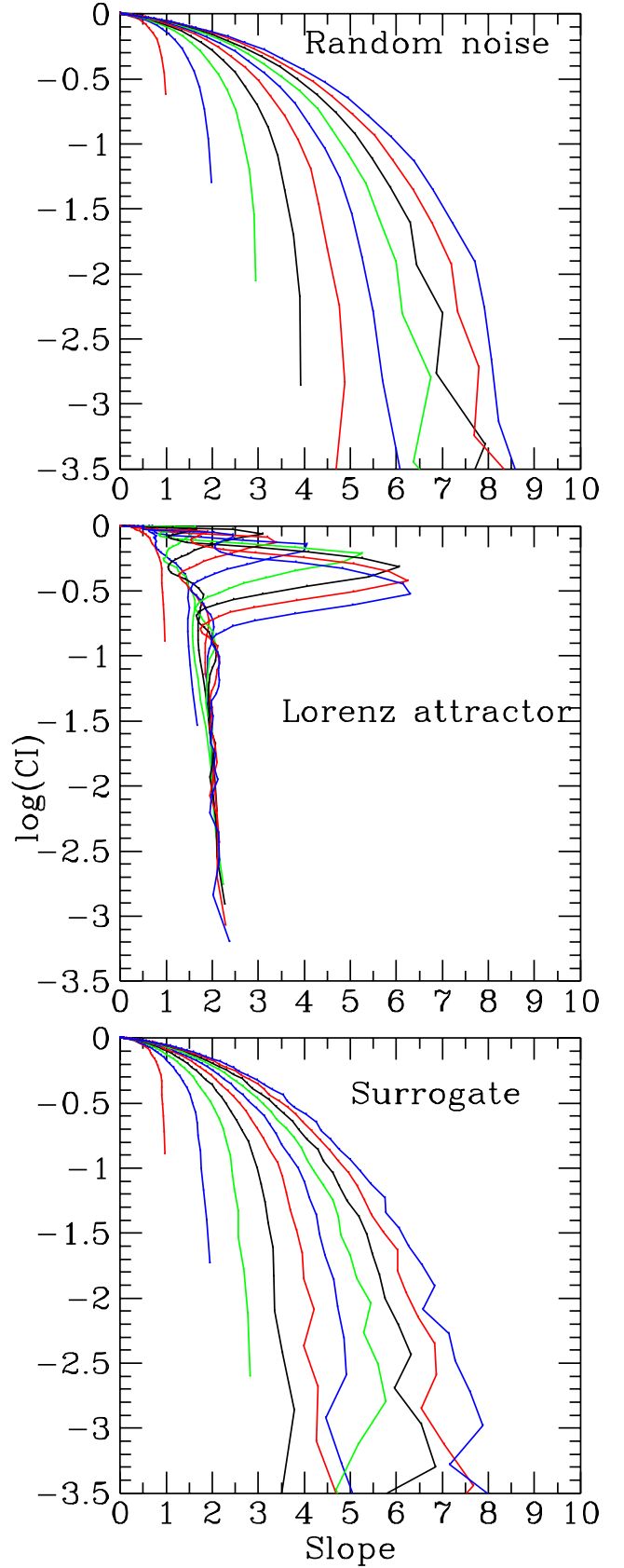,width=8.4truecm,clip=}
}
\caption[h]{
CI diagrams for random noise (upper panel), Lorenz attractor (middle panel) and phase-randomized Lorenz attractor surrogate (lower panel).
For presentation purposes different embedded dimensions, $d=1,2,....10$ (from left
to right) are shown in different colors.
}
\end{figure}

The plan of the paper is the following. In the next section, we describe the data set under
consideration. Subsequently, we discuss the method followed to search for chaos in data and
results in \S3 and present a discussion in \S4. Finally, we end with a summary in \S5.

\subsection[]{Data set}

W2R~1926+42 (19:26:31 +42:09:59, z=0.155) is a BL Lac \citep[an ``LBL",][]{p1}
object that happens to fall in the $Kepler$ satellite field \citep{e1}.
The $Kepler$ continuous monitoring covers a time period of about 1.6~years
(quarters 11 through 17), starting 29.09.2011, with a sampling of about 30~minute and
a duty cycle of about 90\%. The typical photometric errors are less than 0.002\%.
From all available datasets 
we identified 18 uniformly sampled segments of length ranging from about 1300 to
about 16000 points (PDCSAP\_FLUX data, covering roughly 4300-8900\AA{} range).
Occasionally, where some data points (one to rarely upto five) are missing, those are
substituted in terms of linear interpolation, as the CI method requires equidistant
sampling. For longer missing parts, the entire dataset had to be divided into separate
segments. Any remaining parts of length less than of 1000 points (if such) are not considered
in this analysis.

\begin{table*}
\begin{minipage}{126mm}
 \caption{Dataset description}
 \begin{tabular}{ccccc}
\hline
Dataset file name & Segment & Start & End & N \\
\hline
kplr006690887-2012004120508\_llc.fits & seg 1 & 1001.23 & 1032.27 & 1520  \\
                                      & seg 2 & 1033.06 & 1063.22 & 1477  \\
                                      & seg 3 & 1072.56 & 1098.33 & 1262  \\
kplr006690887-2012088054726\_llc.fits & seg 1 & 1126.65 & 1153.99 & 1339  \\
kplr006690887-2012179063303\_llc.fits & seg 1 & 1182.76 & 1214.98 & 1578  \\
                                      & seg 2 & 1215.72 & 1244.92 & 1439  \\
                                      & seg 3 & 1245.86 & 1268.73 & 1120  \\
kplr006690887-2012211050319\_slc.fits & seg 1 & 1274.13 & 1289.47 & 7510  \\
                                      & seg 2 & 1296.40 & 1305.00 & 4210  \\
kplr006690887-2012242122129\_slc.fits & seg 1 & 1306.15 & 1336.31 & 14760 \\
kplr006690887-2012277125453\_llc.fits & seg 1 & 1306.18 & 1336.30 & 1475  \\
                                      & seg 2 & 1337.20 & 1371.32 & 1671  \\
kplr006690887-2012277125453\_slc.fits & seg 1 & 1337.17 & 1371.33 & 16720 \\
kplr006690887-2013011073258\_llc.fits & seg 1 & 1373.51 & 1404.30 & 1508  \\
                                      & seg 2 & 1436.24 & 1471.14 & 1709  \\
kplr006690887-2013098041711\_llc.fits & seg 1 & 1488.69 & 1524.96 & 1776  \\
                                      & seg 2 & 1526.14 & 1557.96 & 1558  \\
kplr006690887-2013131215648\_llc.fits & seg 1 & 1559.25 & 1581.58 & 1094  \\
\hline
 \end{tabular}
\end{minipage}
\end{table*}

Although covered by long cadence datasets (``llc"), the short cadence data (``slc",
sampling of about a minute) are treated separately as they provide a much larger
time resolution and dataset length. For the short cadence datasets where photometric
errors are larger (up to 0.02\%), a median filter is applied over every three data
points. This procedure helps to largely reduce spurious effects of significantly
deviating single points and to reduce almost twice the photometric errors, of course
with the price of reducing the dataset lengths. The data segments used in this analysis are given in Table~1.
The start and end times are in days (BJD-2454833) and N is the number of the points
in the segment.

\subsection{Tests}

The solution of a non-linear dissipative dynamical system is known as $attractor$, if it
stays bound within a finite volume of the phase space \citep[e.g. {\it Lorenz attractor};][]{l2}.
An attractor is $strange$, if it is of non-integer dimension. The
trajectories of a strange attractor evolve in a finite volume of the phase space,
never returning to the same point. The divergence between two infinitesimally close
trajectories increase exponentially in time making the long-term predictions impossible.


The CI method is a suitable approach for revealing signature of low-dimensional chaotic
behavior in time series data \citep{v1}. The method relies on the construction of
a new (empirical) phase space from the available $N$ discrete data points. The data set
(e.g. the light curve) is separated into segments of length $d$. Each segment can be
considered as a $d$-dimensional vector ($X_i$), embedded in the $d$-dimensional empirical
phase space. The number of the vector pairs within a distance smaller than $r$ is computed
as a function of $r$ for different $d$ and related to the total number of pairs ($n_p$)
for that $d$. Thus, the correlation integral can be expressed as
\[C_d(r) = \frac{1}{n_p} \sum_{i,j=1;j\neq i}^{N} \Theta(r-|X_i-X_j|)\]

\noindent where $\Theta$ is the $Heaviside's$ function. Therefore, if the dimension of the attractor is $D$, then
\[ C_d(r) \propto \left\{ 
  \begin{array}{l l}
    r^d,~~  d<D\\
    r^D,~~  d>D.
  \end{array} \right. \]

\noindent Hence, increasing the embedded dimension $d$ leads to saturation when $d>D$, which can be
used for the estimation of the attractor dimension
\[D_c = \lim_{r \to 0} \frac{d{\rm log}C(r)}{d{\rm log}r},\]

\noindent where $D_c$ is the correlation dimension of the attractor and can be a non-integer value.
Knowledge of $D_c$ allows the determination of the number of differential equations, describing
the dynamical system, $N$, which is the first integer value, larger than $D_c$ and therefore makes
possible drawing conclusions about the physical process driving the variability.

The so called Ruelle's criterion \citep{r1} relates the maximal dimension of an attractor (if present),
that can be detected from a time series of $N$ points, with $N$ itself as $D_{max}\leq2{\rm log}N$, which in our case,
depending on the segment, means $D_{max}=$6 -- 8.

As pointed out by \citet{l1}, the results from the CI method might be significantly
contaminated if a strong periodic component ($D=1$) is present in the data. Broadly speaking,
false results of chaotic signatures might be an artifact of the exact shape of the power-density
spectrum (PDS) of the dataset. Therefore, one should also perform the CI test on
``surrogate" data, generated from the original data PDS after phase-randomization
\citep{s1, k1}. If the chaotic signature of the original time series is "real",
such should not be present in the surrogate.

\section{Results}

Before applying the CI method to real data, we test it with light curves of known correlation dimension. Fig. 2 (upper panel) shows
the results for a 1500 points long random noise light curve (infinite dimension). The figure shows the variation of
${\rm log} CI$ with its slope $d{\rm log} CI /d {\rm log} r$, for different embedded dimensions ($d=1..10$), each presented by a
separate curve (of different color in the electronic version). As one sees, each embedded dimension saturates at its corresponding
value, indicating that there is no low-dimensional attractor in these data, as expected. Next (Fig. 2, middle panel) is for a
1500 points long Lorenz attractor light curve ($D_{\rm c}=2.06$). Here all embedded dimensions
of $d>2$ saturate around 2.06, indeed revealing the presence of a low-dimensional attractor. Finally, in the lower panel of Fig. 2,
we see the results for the same Lorenz light curve, this time phase-randomized (surrogate data). As expected, no long dimensional
saturation can be seen, meaning that phase randomization destroys the attractor (provided such is present, which is the case
for a Lorenz attractor).

This simple test suggests that if a low dimensional attractor is really present in the original dataset, it should not be seen in the
phase-randomized surrogates, i.e. both ${\rm log}CI$ versus $d {\rm log} CI / d {\rm log} r$ plots should differ significantly.

Fig. 3 and 4 show CI diagrams for different segments of the blazar light curve. Although the pattern does not resemble the one of the pure
random noise, we cannot find any evidence for saturation (e.g. the presence of a low-dimensional attractor) in any of the segments.
As we have tested only for the first 10 embedded dimensions, one can conclude based on this test that the correlation dimension
(also very similar to so called fractal dimension) could indeed be larger than 10. However, the goal usually is to find
``low" dimensional chaos with $D\sim 3-4$, which may indeed be useful to constrain or develop the theory, as discussed below.
Furthermore, the surrogates show very similar patterns as the real datasets,
meaning that any peculiarity should be attributed to the particular structures in the light curve PDS. Actually, the results are
somewhat similar to those reported by \citet{l1} for their shot-noise model, as well as for most of the sources they studied,
even though the emission mechanisms in these two (blazar and non-blazar) types of objects are very different.

Finding low-dimensional behavior in a dynamical system (in our case -- a light curve) can be of huge importance for the theoretical
models,  as it may limit the number of independent variables (or the number of equations) that governs entirely the system.
People often invoke the so called single zone model
to reproduce the observed blazar SEDs. In this model the synchrotron radiation, which is responsible for the optical emission, is
controlled by only a few parameters - electron density, magnetic field,
Doppler factor, etc. Therefore, if the single zone models actually works most of the time, one should expect to find a low-dimensional
(say D$\simeq$3--4) behavior in blazar light curves.


Indeed, the electron energy density evolution due to external injection and (synchrotron and synchrotron self-Compton, SSC)
radiation loses is described by the {\it Fokker-Planck} (kinetic) equation \citep{k1a}:

\[ \frac{\partial n(\gamma,t)}{\partial t} = \frac{\partial }{\partial \gamma}\left[ |\dot{\gamma}_{tot}|n(\gamma,t) \right] + S(\gamma,t) \]

\noindent where $n(\gamma,t)$ is the relativistic particles energy distribution,
$|\dot{\gamma}_{tot}| = |\dot{\gamma}_{syn}|+|\dot{\gamma}_{ssc}|\propto B^2\gamma^2$ is the total loss and $S(\gamma,t)$ is the
energy injection (source) term.
The synchrotron spectrum (at least its optical part) is entirely determined by $n(\gamma,t)$ \citep[e.g.][Appendix A]{t1}.
In a case of a single monochromatic injection (as frequently assumed) the solution of the kinetic equation is "trivial" in a sense that it cannot
account for the intermittent blazar variability (Fig. 1).

The energy injection (particle acceleration) mechanism is generally unknown, however if the injection occurs near the jet base it might
be considered deterministic in nature, rather than completely stochastic. The reason for this assumption is that in such a case the
accretion flow and/or black hole properties (like spin for instance) should control entirely the relativistic particle acceleration.
Deterministic injection will perhaps lead to deterministic optical variability, evidence for which we did not find from our results.

One way to account for the apparently stochastic light curves, which would be consistent at a larger extent with our findings, is to
invoke a large number of active zones (emitting blobs). The overall light curve in this case will be stochastic in nature, independently
of the exact injection mechanism in each zone.

One such mechanism is based on magnetic reconnections that may occur within the jet and accelerate locally electrons to relativistic energies,
thus leading to intermittent and stochastic in nature variability. The light curve will be perhaps similar to the shot noise case -- i.e.
randomly distributed in time and amplitude explosive events (flares) with possibly different characteristic times. The flares can occur
completely at random or can be a representation of the so called {\it self organized criticality, SOC}, i.e. when one reconnection may lead to
an avalanche of subsequent reconnections \citep[see][for a recent review]{a1}. The latter possibility is supported by the often
reported $1/f$ shape of the blazar PDS's.

Another way to account for the observed variability is to invoke the presence of turbulence in the jet. In these models either turbulent
flow passes through a standing shock \citep[e.g. ][]{m2} leading to instant energy injection in the corresponding (shock crossing)
plasma volume or the opposite -- a shock wave passes though turbulent medium \citep{m1, k2}.
The turbulent flow in these works is modeled as an ensemble of cells with different (randomly distributed within certain limits)
properties, which naturally will lead to stochastic overall variability. This approach, however, might be an oversimplification of turbulence.
Although no fully developed theory of turbulence exists, turbulence and chaos seem to be deeply related \citep{p2}. In fact,
low-dimensional fractal structures have been found in turbulent jets in lab experiments \citep{z1}. So, if turbulence is playing a
major role in shaping blazar jets, one might expect to witness not only fractal spatial structure but perhaps also fractal temporal
behavior (i.e. low-dimensional chaos in the light curves). On the other hand, depending on the exact shock geometry and line of sight angle,
light crossing time delays between emitting cells might be expected \citep{m2}, which may somehow affect the overall light curve
and eventually -- conceal the presence of low-dimensional chaos. But as we stressed above no fully developed turbulence theory is available,
so it is difficult to judge whether a turbulent blazar jet should manifest chaotic temporal behavior or not.

It is interesting to note that there is no practical difference between the CI diagrams for segments covering the high states and the
low states of the light curve. One would normally think that for the low states, there should be less number (perhaps even only one)
of active regions, contributing toward the emission. Should there be many independent regions contributing to the emission it would be
unlikely to find a low dimensional attractor in the light curve. On the other hand, it would be more likely to find such if a single
zone (of a very few) is generating the energy. Not finding a difference in the CI diagrams means there is perhaps no huge difference
in the number zones for low and high states (i.e. being a significant number in both cases). Another explanation, of course, is that
the temporal evolution of a single zone is too stochastic to be described in terms of low-dimensional chaos. Or may be the entire
picture of the blazar emission generation is more complex, than we have previously thought.


\section{Summary}

We have applied the CI method to the longest ever available dataset of a light curve of a $Kepler$ blazar -- W2R 1946+42 to search for
signatures of low-dimensional chaotic behavior. Our results do not reveal such for any of the segments, covering both -- high and
low states of the blazar.

Although not entirely conclusive, these results may be suggestive for the mechanism of particle acceleration. Models, based on a
"single engine", operating perhaps close to the jet base seem unlikely. Models, relying on the presence of many independent active zones,
like multiple magnetic reconnections or turbulence within the jet are plausible.


Further tests are however required for a firm conclusion.

\clearpage

\begin{figure}
{
\psfig{figure=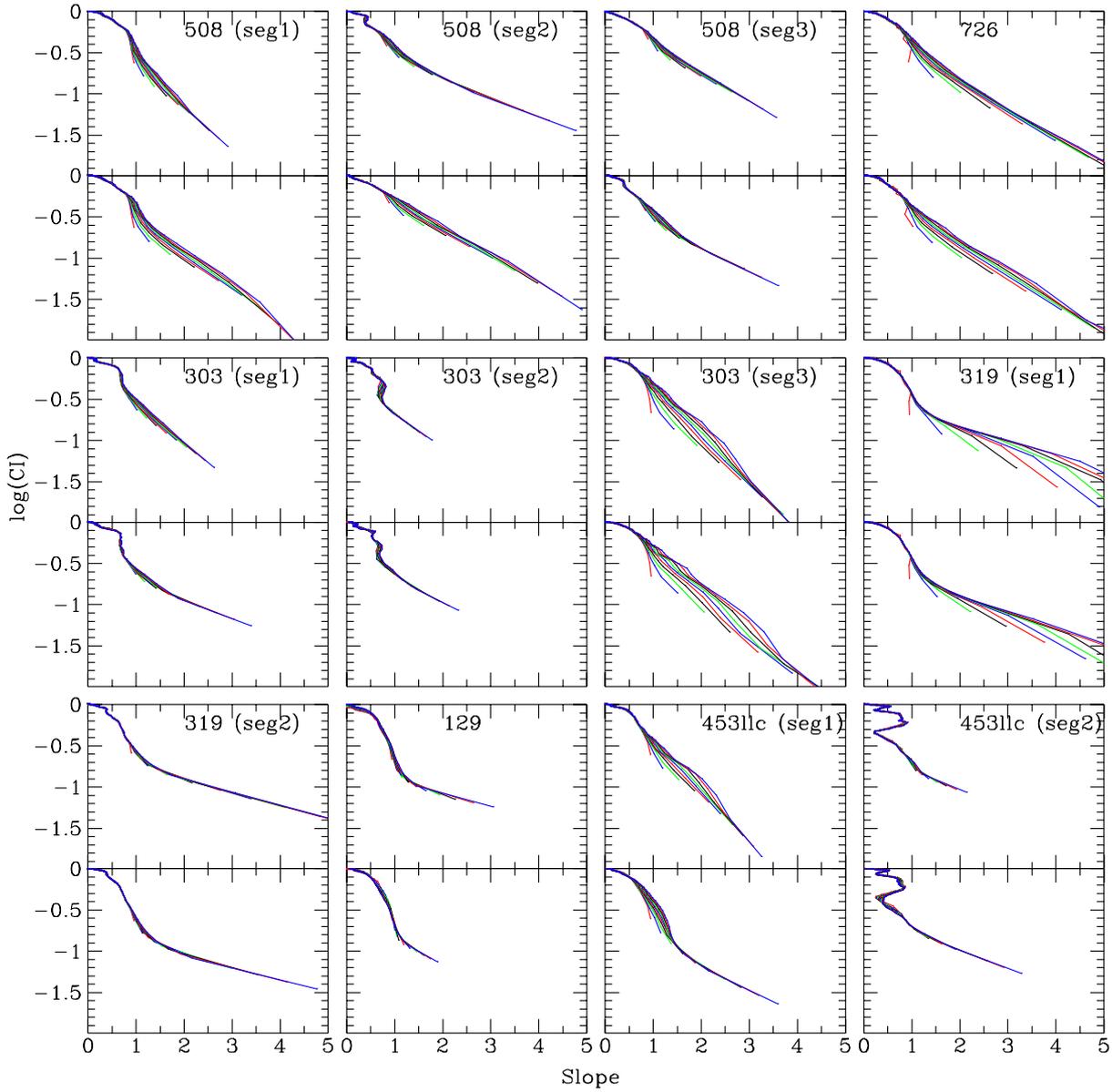,width=16truecm,clip=}
}
\caption[h]{
CI diagrams for different segments of the blazar light curve for
different embedding dimensions, same as in Fig. 2. The last 3 digits of the dataset (see Table 1) are indicated.
The upper panel of each box is the real dataset, the lower one -- the phase-randomized surrogate. None of the diagrams shows clear
indication for a strange attractor of low dimension, especially taking into account that the "surrogate" diagrams are very similar
to the ones of the real data.
}
\end{figure}

\clearpage

\begin{figure}[h]
{
\psfig{figure=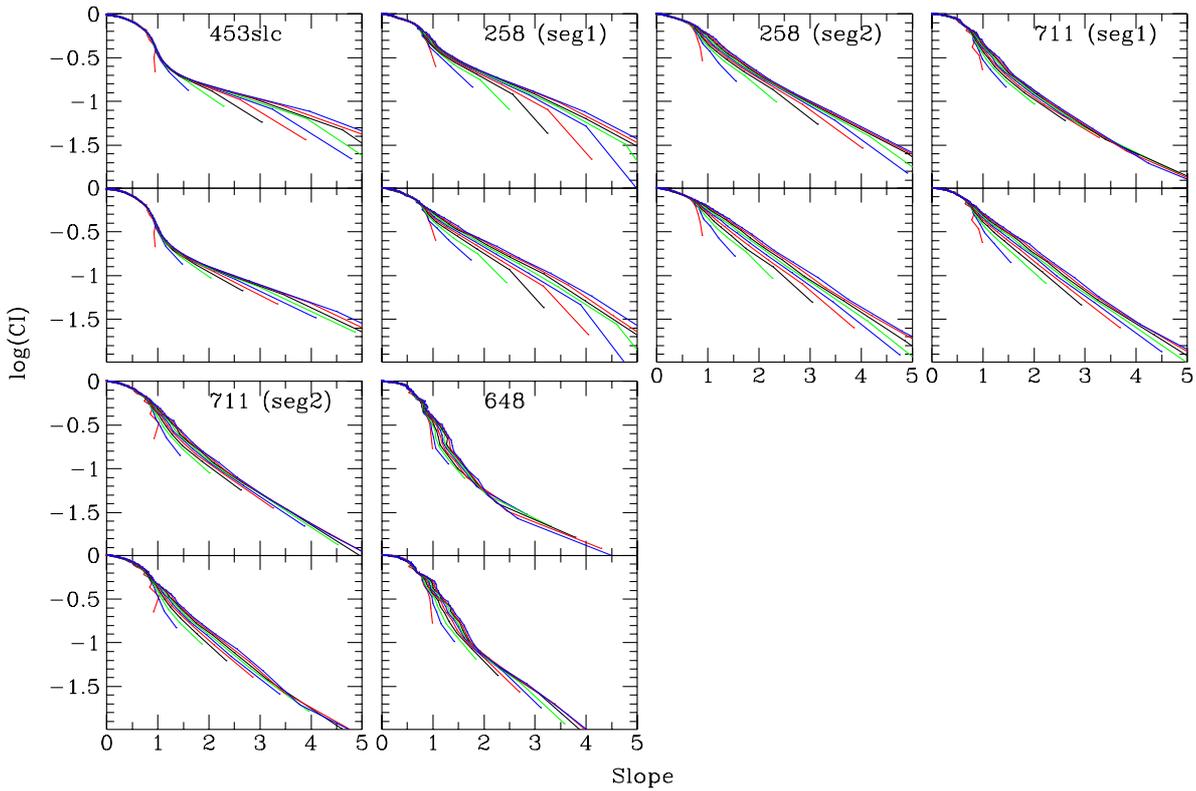,width=16truecm,clip=}
}
\caption[h]{
The same as Fig. 3.
}
\end{figure}

\begin{acknowledgements}
This paper is supported by Bulgarian National Science Fund through grant NTS BIn 01/9 (2013) and
an Indo-Bulgarian Project funded by Department of Science and Technology, India, with grant no.
INT/BULGARIA/P-8/12.
\end{acknowledgements}

\end{document}